\documentstyle[prd,aps,epsfig,eqsecnum,amssymb,amsmath,floats]{revtex}
\begin{document}
\title{Cosmological  constraints from evaporations of primordial black holes.}
\author{
E. V. Bugaev and K. V. Konishchev
}
\address{
Institute for Nuclear Research, Russian Academy of Sciences, 
Moscow 117312, Russia
}
\date{
\today
}
\maketitle
\begin{abstract}
The formula for the initial mass spectrum of primordial black holes (PBHs), which can be  used for a general case of the
scale dependent spectral index, and for a wide class of models of the gravitational collapse, is derived.
The  derivation is based on the Press and Schechter formalism.
The comparative analysis of different types of initial mass spectra used in concrete calculations is carried out.
It is shown that densities of background radiations ($\nu$, $\gamma$)
from PBH evaporations  depend rather strongly on a type of the gravitational collapse
and on a taking into account the spread of horizon masses at which 
PBHs can form. Constraints on parameters of the primordial density perturbation amplitudes based on PBH evaporation 
processes and on atmospheric and solar neutrino data are obtained.
\end{abstract}

\section{Introduction}

Studies of cosmological and astrophysical effects of primordial black holes (PBHs)
are important because they enable one to constrain the spectrum of density fluctuations in the early Universe. If the PBHs
form directly from primordial density fluctuations then they provide a sensitive probe of the primordial power spectrum on small scales,
$\agt 10^{-9} pc$. In particular, limits on PBHs production can be used to constrain models of inflation, in which the perturbation amplitudes are 
relatively large at small and medium scales.

In the simplest case, if we assume that the cosmological PBH formation is dominated, approximately, by primordial perturbations of one 
particular scale (i.e., there is some characteristic epoch of the PBH formation), we can obtain limits on the initial mass fraction of PBHs,
\begin{equation}
\label{int_1}
\rho_i=\rho_{PBH, i}/\rho_{tot, i},
\end{equation}
where $\rho_{PBH, i}$ and $\rho_{tot, i}$ are the PBH and total energy densities, respectively,  at the time $t_i$ of the formation.
This fraction can be expressed by the integral
\begin{equation}
\label{int_2}
\beta_i = \int\limits_{\delta_c}^{1} P(\delta)d\delta ,
\end{equation}
where $P(\delta)$ is a probability distribution for density fluctuations entering horizon at $t_i$, $\delta$ is a density contrast,
and $\delta_c$ is a minimum value of $\delta$ required for the collapse. If the probability distribution  is assumed to be  Gaussian, one has 
\begin{equation}
\label{int_3}
P(\delta)=\frac{1}{\sqrt{2\pi}\sigma} e^{-\frac{\delta^2}{2\sigma^2}}\:,
\end{equation}
where $\sigma$ is the {\it rms} fluctuation amplitude on a given scale. It is just this value  that is determined by the primordial power spectrum.
The connection between $\beta$ and $\sigma$ is very simple in a case of the Gaussian distribution:
\begin{equation}
\label{int_4}
\beta_i\approx \sigma e^{-\frac{\delta_c^2}{2\sigma^2}}.
\end{equation}

The limits on $\beta_i$ arise from the 
entropy production constraints \cite{1}, from a distortion of the microwave background \cite{2}, from 
the cosmological nucleosynthesis constraints \cite{3}. In these cases PBHs which give constraints have 
evaporated completely to the present time. On the contrary,
the gravitational constraint ($\Omega_{PBH, 0}=\rho_{PBH, 0}/\rho_c < 1$),
\begin{equation}
\label{int_5}
\beta_i < 10^{-19}\left(\frac{M}{10^{15}g}\right)^{1/2}\;
\end{equation}
\cite{4}, is valid for PBH masses $M\agt 10^{15} g$ which survive today. 

One should stress that all these limits are based just on the approximation that all PBHs
form at the same scale and, correspondingly, the initial mass spectrum of PBHs is $\delta$-function-like or, at least,  is "non-extended" one.
Only in this case one can  approximately express the observational constraints through the initial mass fraction $\beta_i$.
Evidently,  a more accurate treatment should operate with the initial PBH mass spectrum directly.

The most strong limit on a PBH formation in  the early Universe is due to the possible contribution of
evaporating PBHs to the extragalactic $\gamma$-ray and neutrino backgrounds at energies $\sim 100\text{ MeV}$. 
The limit  of such kind  was obtained in the work of Page and Hawking \cite{5},
authors of which assumed that the differential initial mass spectrum of PBHs has power law form predicted in Carr's work \cite{6},
  
\begin{equation}
\label{int_6}
n_{BH}(M_{BH})=(\alpha -2)\left(\frac{M_{BH}}{M_{*}}\right)^{-\alpha}M_{*}^{-2}\rho_c\Omega_{PBH, 0} \;,
\end{equation}
$$
M_{*}\alt M_{BH}.
$$
Here, $\alpha=2.5$ for PBHs formed in radiation-dominated era, $M_{*}$ is the mass of a black hole whose life-time is equal to the present 
age of the Universe.

This PBH mass spectrum is, clearly, the example of  an "extended" spectrum (it was derived in \cite{6} by considering the PBH formation as a process 
stretched in time). Naturally, the famous Page-Hawking constraint was formulated in terms of a PBH number density rather than  in terms of $\beta_i$. According
to \cite{5}, the upper limit on the present PBH number density is $\sim 10^{4}pc^{-3}$ (or, that  is the same, $\Omega_{PBH, 0}\alt 10^{-8}$). This constraint was
improved in later works \cite{7}, where the same initial PBH mass spectrum was used as input.

The derivation of Eq.(\ref{int_6}) was based on the assumption of exact scale invariance (scale-independence of the perturbation amplitudes).
The Carr's work \cite{6} appeared before an advent of the inflation hypothesis, and at that time it seemed improbable that the possible case of a growth
of the perturbation amplitudes with a decrease of the scale can be of any importance (from the point of view of observational evidences of the
PBHs existence). Now we know that, due to inflation, minimum values of PBH mass in the initial mass spectrum can be rather large ($10^{13}-10^{14}$~g
or even more). It means that the PBHs can have  rather extended mass spectrum even if the primordial fluctuations are not strictly scale invariant.

In general, initial PBH mass spectrum depends on cosmological and astrophysical aspects of the model used for its derivation
(and, through the model, on such parameters as a time of the end of inflation, a reheating temperature, a spectral index of the density perturbations 
(or parameters of the inflationary potential)
and, last but not least,on  parameters, characterizing the process of the gravitational collapse leading to the PBH's birth). Evidently, observational
constraints on the PBH production (especially those derived from
measurements of extragalactic diffuse backgrounds of $\gamma$-rays and neutrino) can be used as constraints on at least one of these parameters.
Clearly, these constraints  depend on the initial PBH mass spectrum, i.e., on the parameters of 
this spectrum which are considered as free in a course of the constraint's derivation.

Usually, the parameters characterizing primordial density fluctuation amplitudes (e.g., the spectral index, if it  is scale independent)
are objects of the constraining. Parameters of the gravitational  collapse and a method used for a summation over epochs of the PBH  formation
(if such a summation is performed) determine, by definition, the {\it  type} of the  initial PBH  mass spectrum. All  other parameters (characterizing, in particular,
the end of inflation and the beginning of radiation dominated era) are "external" and  considered as free  ones.

In the present work we obtain some cosmological constraints following from the possible contribution of PBH evaporations in extragalactic diffuse
neutrino background. Assuming that the power spectrum of primordial fluctuations has a power-law form, $P(k)\sim k^n$, we present the
observational limits as constraints on values of the spectral index $n$ (using the normalization on COBE data). In a more general case of the non-power 
$P(k)$-dependence one can directly constrain parameters of a concrete inflation model. We consider, as an example, a case of the running mass
inflation model \cite{8} and obtain constraints on parameters of the corresponding inflationary potential.

The main physical assumption used in the work is that a PBH formation process occurs during radiation dominated era only.
The main "external" parameter is $t_i$, a time of the beginning of radiation era. At this moment of time we have,
by assumption, the primordial spectrum of
density perturbations, and no primordial black holes. In a short period between the end of inflation and $t_i$
there can be additional amplification of density perturbation amplitudes (e.g., in a preheating phase)
but, by assumption, in this period there is no PBH formation. The time $t_i$ is connected by a usual way with a corresponding initial 
temperature of radiation era,  which we call a reheating temperature $T_{RH}$.

The paper is organized as follows. In Sec.\ref{sec:Init_mass_sp} 
we derive, using the Press and Schechter formalism, the general 
formula for the initial PBH spectrum which is valid for any law of 
$P(k)$-dependence and for a wide class of models of the gravitational 
collapse. In Sec.\ref{sec:Diff_typ_sp} we give  the comparative
analysis of different types of initial PBH mass spectra used in concrete calculations.
In Sec.\ref{sec:Neut_bg_sp} the neutrino background spectra from PBH evaporations 
are derived and the corresponding constraints on cosmological parameters are obtained.
Discussions and conclusions are presented in Sec.\ref{sec:Dis_and_Con}.

\section{Initial PBH mass spectrum formula}
\label{sec:Init_mass_sp}

According to the Press - Schechter theory \cite{9}, the mass distribution function
$n(M,\delta_c)$, which is defined such that $n(M,\delta_c)dM$ 
is the comoving number density of gravitationally bound objects
in the mass range $(M,M+dM)$, is determined by the equations 
(see, e.g., \cite{10})

\begin{equation}
\label{1_1}
n(M,\delta_c)=\frac{\rho_i}{M}\left|\frac{\partial F}{\partial M}(M,\delta_c)\right|,
\end{equation}

\begin{equation}
\label{1_2}
F (M,\delta_c)=\sqrt{\frac{2}{\pi}}\frac{1}{\sigma_R (M)} e^{-\frac{\delta^2}{2\sigma_R^2 (M)}}\;. 
\end{equation}
Here, $\sigma_R (M)$ is the {\it rms} mass fluctuation on a mass scale $M$, i.e., 
a standard deviation of the density contrast of the density  field smoothed on a size  $R$  having mean mass $M$,
$F (M,\delta_c)$ is a fraction of the volume occupied by the regions
that will eventually collapse into bound objects with masses larger than $M$ ($\delta_c$ is a minimum value of the density
contrast needed for developing a nonlinear growth of the density fluctuations
in the overdense region).
The density $\rho_i$ is, in our case, the background energy density at the initial moment of time, $t=t_i$ 
(it is a moment of a beginning of the growth of the density fluctuations).

By the construction, Eq.(\ref{1_1}) supposes a step-by-step formation of the  whole mass spectrum of
gravitationaly bound objects (PBHs,  in our case). Heavier PBHs form at later times when
corresponding masses of overdense regions cross horizon. The nonzero probability that small
PBHs are included in larger ones at later times is approximately taken into account
by the differentiation of $F(M,\delta_c)$ with $M$. The correct normalization of the spectrum
is ensured by the extra factor 2 in Eq.(\ref{1_2}).    

Introducing the double differential distribution $n(M,\delta)$ \cite{11} defined by
the expression
\begin{equation}
\label{1_3}
n(M,\delta_c)=\int\limits_{\delta_c}^{\infty} n(M,\delta)d\delta ,
\end{equation}
one obtains from Eqs.(\ref{1_1}-\ref{1_2})
\begin{equation}
\label{1_4}
n(M,\delta)=\sqrt{\frac{2}{\pi}}\frac{\rho_i}{M}\frac{1}{\sigma_R^2 (M)}
\left|\frac{\partial \sigma_R (M)}{\partial M}\left(\frac{\delta^2}{\sigma_R^2 (M)}-1\right)\right|
e^{-\frac{\delta^2}{2\sigma^2_R (M)}}.
\end{equation}
The {\it rms} mass fluctuation $\sigma_R (M)$ is given by the integral \cite{12}
\begin{equation}
\label{1_5}
\sigma_R^2 (M)=\int \left(\frac{k}{aH}\right)^4 \delta_H^2 (k) W^2(kR) T^2(k)\frac{dk}{k},
\end{equation}
where $W(kR)$ is the smoothing window function, $\delta_H (k)$ is the (nonsmoothed)
horizon crossing amplitude, $T(k)$ is the transfer  function taking into account the 
microphysical processing of  density perturbations after entering horizon, $a$ and $H$ are a cosmic
scale factor and  an expansion rate.

There are two time-dependent factors in the right hand side of  Eq.(\ref{1_5}):
the factor $(aH)^4$  in the denominator describing a growth of the perturbations with  time
and the transfer function. We need the amplitude $\sigma_R (M)$ at the initial moment  of time,
$t=t_i$. A role of the transfer function is  essential at  large $k$, if 
an integrand in Eq.(\ref{1_5}) is not cut off by a $k$-dependence of the window function
and $\delta_H(k)$. In  the case when a top-hat window function is used  and $\delta_H^2\sim k^{n-1}$
with $n>1$, the contribution of large   $k$ modes in the integrand  is rather large and a convergence
of the integral  in Eq.(\ref{1_5}) is  provided just by the transfer function. 
The  form of this function at an epoch of the PBH formation is, of course, model dependent.
One can assume, for simplicity, that  at
$t=t_i$ the transfer function has a simplest form:
\begin{equation}
\label{1_5a}
\left. T(k) \right| {}_{t=t_i} =\Theta (k_e-k),
\end{equation}
where $\Theta (x)$ is the step function, and the border value $k_e$ is of the order of $k_{end}$, the horizon scale 
at the end of inflation \cite{13}.  From our point of  view, however, one cannot exclude the possibility 
that $k_e >> k_{end}$.
If Eq.(\ref{1_5a}) is assumed, the equation (\ref{1_5}) can be rewritten, for a power-law of the primordial power spectrum and
a top-hat window function, as 
\begin{equation}
\label{1_5b}
\sigma_R^2 (M) =\frac{k_{fl}^4}{(aH)^4}\cdot C^2\left(\frac{k_e}{k_{end}},n\right)\cdot\delta_H^2(k_{fl})\;\;,
\end{equation}
\begin{equation}
\label{1_5c}
C^2\left(\frac{k_{e}}{k_{end}},n\right)=\int\limits_0^{k_e/k_{end}} x^{n+2} W^2 (x) d x = 9\!\!\!\!\!\int\limits_0^{k_e/k_{end}} x^{n+2}\left[\frac{\sin x}{x^3}
-\frac{\cos x}{x^2}\right]^2 d x.
\end{equation}

If the spectral index $n$ is slightly more than one, $n\cong 1.2$, and if $k_{e}/k_{end}\cong 10$ one has $C\sim 5$,
i.e., the value  of $C$ is close to the corresponding value \cite{4} for the present epoch. Having no reliable model for the transfer  function at
early postinflationary epoch,  we assume  everywhere below,  where 
a power-law  form of  $P(k)$ is used, that $C$ is constant, and equal to $4.7$, throughout all times. Correspondingly,
\begin{equation}
\label{1_6}
\sigma_R^2 (M)\approx  \frac{k_{fl}^4}{(aH)^4}C^2\delta_H^2(k_{fl})= \frac{k_{fl}^4}{(aH)^4}
\sigma_H^2 (M_h)
\;.
\end{equation}

In Eq.(\ref{1_6}) $k_{fl}$ is the comoving wave number, characterizing the perturbed region,
\begin{equation}
\label{1_7}
k_{fl}=\frac{1}{R},
\end{equation}
and $M_h$ is  the fluctuation mass at the moment when the perturbed
region crosses horizon in radiation dominated era. The value of $M_h$ is expressed through $k_{fl}$ (see below). 
The function $\sigma_H (M_h)$ introduced by Eq.(\ref{1_6}) is, evidently, the smoothed horizon crossing 
amplitude (a standard deviation of the density contrast at a moment when the fluctuation crosses horizon).

The general formula for $\sigma_H^2 (M_h)$, which is valid for an arbitrary $k$-dependence of $P(k)$, is 
\begin{equation}
\label{1_7a}
\sigma_H^2 (M_h)=\int\limits_0^{k_e/k_{end}} x^3 W^2 (x)\delta_H^2 (k_{fl} x) d x.
\end{equation} 

At the moment $t_h$ when the fluctuation crosses horizon one has
\begin{equation}
\label{1_8}
R a(t_h)\approx t_h,
\end{equation}
and, in radiation dominated era $(a\sim\sqrt{t})$, 
\begin{equation}
\label{1_9}
M_h\sim\frac{1}{k_{fl}^2}.
\end{equation}

At the same time, an initial value of the fluctuation mass, $M$, is proportional to $R^3$,
so $M_h\sim M^{2/3}$. The exact connection is
\begin{equation}
\label{1_10}
M_h=M_i^{1/3} M^{2/3}\;,
\end{equation}
where $M_i$ is the horizon mass at the initial moment of time, $t=t_i$,

\begin{equation}
\label{1_11}
M_i\sim t_i\sim \frac{1}{(a_i H_i)^2}\;.
\end{equation}

The denominator $(aH)^4$ in Eq.(\ref{1_6}) must be taken also at $t=t_i$ and,
using the proportionality
\begin{equation}
\label{1_12}
M\sim R^3a^3(t_i)\rho(t_i)\sim \frac{R^3}{\sqrt{t_i}}\sim k_{fl}^{-3}\frac{1}{\sqrt{t_i}},
\end{equation}
one can easily see that
\begin{equation}
\label{1_13}
\frac{k_{fl}^2}{(a_iH_i)^2}=\left(\frac{M}{M_i}\right)^{-2/3}.
\end{equation}
Due to Eq.(\ref{1_13}), one has the simple connection between $\sigma_R$ and  $\sigma_H$:
\begin{equation}
\label{1_14}
\sigma_R (M)=\left(\frac{M}{M_i}\right)^{-2/3}\sigma_H (M_h).
\end{equation}
For the following we parametrize the $k$-dependence of  $\delta_H(k)$ by the formula
\begin{equation}
\label{1_15}
\delta_H(k)=\delta_H(k_0)\left(\frac{k}{k_0}\right)^{\frac{n(k)-1}{2}}.
\end{equation}
Introducing now the new variable $\delta'$, 
\begin{equation}
\label{1_16}
\delta'=\delta\left(\frac{M}{M_i}\right)^{2/3}
,
\end{equation}
and a general form of the connection between  PBH mass, fluctuation mass and density
contrast,
\begin{equation}
\label{1_17}
M_{BH}=f(M,\delta'),
\end{equation}
one gets the PBH mass spectrum $n_{BH} (M_{BH})$:
\begin{equation}
\label{1_18}
n_{BH}(M_{BH})=\int n_{BH} (M_{BH},\delta')d\delta'\;,
\end{equation}
\begin{equation}
\label{1_19}
n_{BH}(M_{BH},\delta')=n(M,\delta)\frac{d\delta}{d\delta'}\frac{dM}{dM_{BH}}.
\end{equation}

Using Eqs.(\ref{1_4}),(\ref{1_6}),(\ref{1_10}),(\ref{1_14}-\ref{1_17}), one obtains
the final expression
 for  the PBH mass spectrum
\begin{eqnarray}
\label{1_20}
n_{BH}(M_{BH})=\sqrt{\frac{2}{\pi}}{\rho_i}\int\frac{1}{M\sigma_H  (M_h)}
\left\{
\frac{2}{3M}-\left[
\frac{n'}{2}ln \frac{k_{fl}}{k_0}+\frac{n(k_{fl})-1}{2}\frac{1}{k_{fl}}
\right] \cdot\frac{\partial k_{fl}}{\partial M}
\right\}\times
\nonumber\\
\\
\left|\frac{\delta'{}^2}{\sigma_H^2 (M_h)}-1\right|
e^{-\frac{\delta'{}^2}{2 \sigma_H^2 (M_h) }}\frac{d\delta'}{d f(M,\delta')/dM}\;.
\nonumber
\end{eqnarray}

Here  the following notation is used: 
\begin{equation}
n' = \left. \frac{d n(k) } {d k}  \right|_{k=k_{fl}}.
\end{equation}
The limits of integration in Eq.(\ref{1_20}) are determined from Eqs.(\ref{1_16}) and (\ref{1_17}) using the conditions
\begin{equation}
\delta_{min}=\delta_c\;\;\;,\;\;\; M_{min}=M_{h}^{min}=M_i \;\;\;.
\end{equation}

The connection between $k_{fl}$ and $M$ is  determined by the expressions

\begin{eqnarray}
\label{1_27}
k_{fl}=(a_{eq}H_{eq})\left(\frac{M_{eq}}{M_h}\right)^{1/2},\;\;\;\;\;\;\;\;\;\;\;\;\nonumber\\
\\
a_{eq}H_{eq}=\sqrt{2}H_0 \Omega_m \Omega_r^{-1/2} \;\;\;,\;\;\; M_{eq}=\frac{1}{8}\frac{M_{pl}}{t_{pl}}t_{eq}\nonumber
\end{eqnarray}
together with Eq.(\ref{1_10}). In these formulae $a_{eq}$ and $H_{eq}$ are the cosmic scale  factor and the   expansion rate
at the  moment of matter-radiation equality, $t=t_{eq}$, $H_0$ is the Hubble constant ($H_0=100 h km s^{-1} Mpc^{-1}$),
$M_{eq}$ is the horizon mass at $t_{eq}$, $\Omega_m$ and $\Omega_r$ are matter and radiation densities in
units of $\rho_c$. We consider in this paper,  for simplicity, the case  of flat models with zero cosmological
constant.

\section{Different types of initial PBH mass spectra}
\label{sec:Diff_typ_sp}

A concrete form of the initial mass spectrum of primordial black holes is determined, first of all, by
peculiarities of the gravitational collapse near a threshold of the black
hole formation. According to analytic calculations of seventies \cite{15,16}
a critical size of the density contrast needed for the PBH formation, $\delta_c$,
is about $1/3$. Besides, it was argued that all PBHs have mass roughly
equal to the horizon mass at a moment of the formation, independently
of the perturbation size. The approximate connection between the PBH mass
and the horizon mass in such models is very simple \cite{18,11}:
\begin{equation}
\label{0}
M_{BH}=\gamma^{1/2} M_{h},
\end{equation}
where $\gamma$ is the ratio  of  the  pressure to energy density ($\gamma=1/3$  in radiation dominated era).

However, it was shown recently
that near a threshold of the black hole formation the gravitational collapse
behaves as a critical phenomenon \cite{16}. In this case the initial mass function
will be quite different from the analogous function in the analytic calculations \cite{15}.
The main feature is that the PBH mass formed depends on a size of the
fluctuation \cite{16},
\begin{equation}
\label{1}
M_{BH}=\kappa M_h (\delta'-\delta_c)^{\gamma_k}.
\end{equation}
In this formula $M_h$ is the horizon mass at the time when the fluctuation
enters horizon; $\delta_c$, $\kappa$, $\gamma$ are parameters of a concrete
model of the critical collapse \cite{16,17}. The corresponding $f(M,\delta')$ function
defined in Eq.(\ref{1_17}) is determined using the  relation (\ref{1_10}):
\begin{equation}
f(M,\delta')=\kappa M_i^{1/3} M^{2/3} (\delta'-\delta_c)^{\gamma_k}.
\end{equation}
It is seen from Eq.(\ref{1})
that the PBH mass may be arbitrarily small independently of the value of $M_h$.
Besides, a value of the critical overdensity, $\delta_c$, in such models
is typically $\sim 0.7$, i.e., about a factor of 2 larger than the value used in the 
standard collapse case.

The second important ingredient of a PBH initial mass spectrum calculation
is a taking into account the spread of horizon masses at which PBHs are formed. 
This problem exists independently of a nature of the gravitational collapse.  

As is stated in the {\it Introduction},
it is assumed very often that a majority of the PBH formation processes occurs
at the shortest possible scale. In particular, the authors of ref.\cite{16}
determined the PBH initial mass function under the assumption
that all PBHs form at the same horizon mass. The accuracy of such an approximation
was studied in the work \cite{19} using the excursion set formalism, 
and it was found that it is good enough, at least in a case of  the power-law
density perturbation spectra with the spectral index close to $1.2$ (i.e.,
slightly "blue" spectra, with the value of $n$ satisfying observational constraints).

As it follows from the previous Section, the effect of an  accumulation 
of PBHs formed at all epochs after the beginning of ordinary radiation era
can be conveniently taken into account by the Press-Schechter formalism  \cite{9}.
For a particular  case of the critical collapse, the expression for the PBH  mass
spectrum, based on the Press-Schechter formalism, was obtained in refs.\cite{20,21}.
The spectrum formula  is \cite{21}

\begin{equation}
\label{4}
n_{BH}(M_{BH})=\frac{n+3}{4}\sqrt{\frac{2}{\pi}}\rho_i M_i^{1/2} M_{BH}^{-5/2}\int
\limits_{\delta_c}^{\delta_{max}}\frac{1}{\sigma_H(M_h)} \left|\frac{\delta'{}^2}{\sigma_H^2(M_h)}-1\right|
e^{-\frac{\delta'{}^2}{2\sigma_H^2(M_h)}}\xi^{3/2} d\delta',
\end{equation}
$$
\xi\equiv \kappa(\delta'-\delta_c)^{\gamma_k}.
$$
An upper limit of 
the integration in Eq.(\ref{4}) is determined by the expression
\begin{equation}
\label{5}
\delta_{max}=min\left[\left(\frac{M_{BH}}{\kappa M_{i}}\right)^{1/{\gamma_k}}\!\!\!\!+\delta_c,\,\,\,1\right].
\end{equation}

The formula (\ref{4}) was derived using the assumption that the primordial power spectrum $P(k)$ has
a power-law form  (i.e., the spectral index $n$ is constant throughout all scales). Correspondingly,
Eq.(\ref{4}) can be obtained from the general expression (\ref{1_20}) omitting there  the $n'$-term.  
The amplitude $\sigma_H (M_h)$ can be normalized on COBE data  using the 
connection between $\sigma_H$ and $\delta_H$ given by Eq.(\ref{1_6}), 
\begin{equation}
\label{2}
\sigma_H(M_h)=C\delta_H(k_{fl}),
\end{equation}
and the relation \cite{4,11}
\begin{equation}
\label{3}
\delta_H(k_{fl})\cong 2\cdot 10^{-5} \left(\frac{M_{eq}}{M_{h\!0}}\right)^{\frac{1-n}{6}}
\left(\frac{M_h}{M_{eq}}\right)^{\frac{1-n}{4}}.
\end{equation}
Here $M_{h\!0}$ is the present horizon mass.

All numerical calculations  the results of which are presented in this paper
are carried out with the following values of the parameters \cite{16}:
\begin{equation}
\label{6a}
\kappa=3\;\;\;,\;\;\; \gamma_k=0.36\;\;\;,\;\;\; \delta_c=0.7.
\end{equation}

The known formula \cite{18} for the PBH mass spectrum in the  standard collapse case can be obtained from Eq.(\ref{4}) by
using the substitutions \cite{21}
\begin{equation}
\label{6b}
\gamma_k \to 0 \;\;\;,\;\;\; \kappa\to \gamma^{1/2}\;\;\;,\;\;\; \delta_c \to \gamma
\end{equation}
and the approximate relation 
\begin{equation}
\label{6c}
\int^1_{\gamma} d\delta ' \left(\frac{\delta '{}^2}{\sigma_H^2} -1 \right) e^{-\frac{\delta '{}^2}{2\sigma_H^2}}
\approx \gamma e^{-\frac{\gamma^2}{2\sigma_H^2}}\;\;.
\end{equation}
Substituting Eqs.(\ref{6b}) and (\ref{6c}) in Eq.(\ref{4}) one  obtains
\begin{equation}
\label{6d}
n_{BH}(M_{BH})=\frac{n+3}{4}\sqrt{\frac{2}{\pi}}\gamma^{7/4}\rho_iM_i^{1/2}
M_{BH}^{-5/2}\sigma_H^{-1}(M_h)\exp\left(-\frac{\gamma^2}{2\sigma_H^2(M_h)}\right)\;.
\end{equation}

The   Carr's formula [Eq.(\ref{int_6})] follows from Eq.(\ref{6d}) if two additional assumptions are used: 
i)   the amplitude $\sigma_H(M_h)$ does not depend on $M_h$ (exact scale invariance) and 
ii)  there is no minimum value of PBH mass in the initial spectrum. 
In this case the spectrum is $AM_{BH}^{-5/2}$, as it follows from Eq.(\ref{6d}).
The factor $A$ is found by normalization of the present PBH
mass spectrum on the present value of $\rho_{PBH, 0}=\Omega_{PBH, 0}\rho_c$, using the approximate formula
for the instantaneous PBH spectrum at $t=t_0$ \cite{11},
\begin{equation}
\label{6e}
n_{BH}(m,t_0)=\frac{m^2}{(M_*^3 + m^3)^{2/3}} A (M_*^3 + m^3)^{-5/2}\Theta (m-M_*),
\end{equation}
($\Theta (x)$ is the step function), and the expression
\begin{equation}
\label{6f}
\rho_{PBH, 0}=\int\limits_{M_*}^{\infty} m n_{BH} (m,t_0) d m.
\end{equation}

We  carried out in this work the calculations of the neutrino background spectra from PBH
evaporations using the mass spectra given by Eqs.(\ref{4}),(\ref{6d}) and, in parallel, the PBH mass spectrum
from ref.\cite{16},
\begin{equation}
\label{6}
n_{BH}(M_{BH})=\frac{\rho_i\left(\frac{M_{BH}}{k M_i}\right)^{1/{\gamma_k}}}{\sqrt{2\pi}
\sigma_H(M_h) M_{BH} M_i \gamma_k} e^{-\frac{\left[\left(\frac{M_{BH}}{k M_i}\right)^{1/{\gamma_k}}
+\delta_c\right]^2}{2\sigma_H^2(M_h)}}.
\end{equation}
In this expression the amplitude $\sigma_H$ is determined by the same formula (\ref{2}),
as above, but with the constant value of $M_h$, $M_h=M_i$, in accordance with the assumption
of authors of ref.\cite{16} that all PBHs form at the smallest scale.  

We have, for calculations of $n_{BH}(M_{BH})$, using Eqs.(\ref{4}),(\ref{6d}) and (\ref{6}), 
two free parameters: the spectral index $n$, giving the perturbation amplitude 
through the normalization on COBE data on large scales, and $t_i$, the moment of time just after reheating, from
which the process of PBH formation started. The value of $t_i$ is connected, in our approach, with a value of the
reheating temperature,
\begin{equation}
\label{7}
t_i=0.301 g_*^{-1/2}\frac{M_{pl}}{T_{RH}^2}
\end{equation}
($g_*\sim 100$ is a number of degrees of freedom in the early Universe).
The  initial horizon mass $M_i$ is expressed through~$t_i$:
\begin{equation}
\label{7a}
M_i\cong \frac{4}{3}\pi  t_i^3\rho_i.
\end{equation}

Typical results of PBH mass spectrum calculations, for the case of a scale independent spectral index, are shown on Fig.1.
We use, for convenience, the following abbreviations: KL, NJ and BK signify, correspondingly,
the PBH mass spectra calculated by formulae of Kim and Lee \cite{18} (Eq.(\ref{6d})), Niemeyer and Jedamzik \cite{16} (Eq.(\ref{6}))
and Bugaev and Konishchev \cite{11} (Eq.(\ref{4})).
One can see from Fig.1 that at large values of the spectral index ($n=1.3$) there is a clear difference
in a behavior of two mass spectra, NJ and BK, based both on a picture of the critical collapse:
BK spectrum is less steep at $M_{BH}>M_{i}$. Of  course, this difference  is not so spectacular in cases when
the spectral index $n$ is more close to 1.

The common feature of NJ and BK spectra is the maximum at $M_{BH}\cong M_i$. Besides, both spectra have the same slope
in the region $M_{BH}<M_i$. The distinctive peculiarity of KL spectra is a sharp cut off at $M_{BH}=M_i$, due to
the rigid connection between PBH mass and horizon mass [Eq.(\ref{0})] in the standard collapse case.

\section{Neutrino background spectra from PBHs evaporations}
\label{sec:Neut_bg_sp}

Evolution of a PBH mass spectrum due to the evaporation leads to the
approximate expression for this spectrum at any moment of time:
\begin{equation}
\label{8}
n_{BH}(m,t)=\frac{m^2}{(3\alpha t + m^3)^{2/3}} n_{BH} \left((3\alpha t + m^3)^{1/3}\right),
\end{equation}
where $\alpha$ accounts for the degrees of freedom of evaporated particles and, strictly 
speaking, is a function of a running value of the PBH mass $m$. In all our numerical
calculations we use the approximation
\begin{equation}
\label{9}
\alpha=const=\alpha (M_{BH}^{max}),
\end{equation}
where $M_{BH}^{max}$ is the value of $M_{BH}$ in the initial mass spectrum
corresponding to a maximum of this spectrum. Special study shows that errors
connected with such approximation are rather small.

The expression for a spectrum of the background  radiation is \cite{11}
\begin{eqnarray}
\label{10}
S(E)=\frac{c}{4\pi}\int dt \frac{a_0}{a}\left(\frac{a_i}{a_0}\right)^{3}
\int dm \frac{m^2}{(3\alpha t + m^3)^{2/3}} n_{BH} \left[(3\alpha t + m^3)^{1/3}\right]
\cdot f(E\cdot (1+z),m)e^{-\tau(E,z)}\nonumber\\
\\
\equiv \int F(E,z)d \log_{10} (z+1).\nonumber
\end{eqnarray}

In this formula 
$a_i$, $a$ and $a_0$ are cosmic scale factors at $t_i$, $t$ and at   present time,
respectively, and
$f(E,m)$ is a total instantaneous spectrum of the background radiation
(neutrinos or photons) from the black hole evaporation. It includes the pure Hawking 
term and contributions from fragmentations of evaporated quarks and from decays of
pions and muons (see \cite{11} and  earlier papers \cite{22,23,24} for details). 
The exponential factor in Eq.(\ref{10})
takes into account an absorption of the radiation during its propagation in
space. The processes of the neutrino absorption are considered, in a given context,
in ref.\cite{11}. 

In the last line of Eq.(\ref{10}) we  changed the variable $t$ on $z$ using
the flat model with $\Omega_{\Lambda}=0$
 for which
\begin{eqnarray}
\label{11}
\frac{dt}{dz}=-\frac{1}{H_0 (1+z)}\left(\Omega_m (z+1)^{3}+\Omega_r (z+1)^{4}\right)
^{-1/2},\nonumber\\
\\
\Omega_r = (2.4\cdot 10^{4} h^2)^{-1} \;\;\;,\;\;\; h=0.67.\nonumber
\end{eqnarray}

On first five figures (Figs.\ref{fig:fig2}-\ref{fig:fig6}) some results of calculations for the case
of a scale independent spectral index are presented.
Several examples of $z$-distributions (the integrands of the integral over $z$ in Eq.(\ref{10}))
are shown on Fig.2. Again one can see the rather strong difference of two cases: the 
existence of a tail of heavy masses in BK spectrum leads to a relative enhancement
of low $z$-contributions. The effect becomes more distinct with a rise of the reheating
temperature.
 
The sharp cut-off of all $z$-distributions near $z\sim 10^7$ is entirely due to the neutrino
absorption 
\cite{11}. The shrinkage of $z$-distributions at large $T_{RH}$ in NJ-case is due to the absence
of large masses in the spectrum (PBHs of small masses evaporated earlier, and their radiation today is more redshifted).

Some results of  calculations of the neutrino background spectra are shown on Figs.3-5.
The functional form of these spectra is qualitatively the same as in the standard collapse case \cite{11}:
$E^{-3}$ law at large neutrino energies and a flat part at low energies with the crossover energy
depending on $T_{RH}$. The only new feature is the sharp steepening of the spectra at high values
of $T_{RH}$ in NJ case (Fig.4) connected with the corresponding shrinkage of  the $z$-range.

Fig.5 shows the sensitivity of background neutrino intensities to a chosen value of the spectral index~$n$,
at a fixed $T_{RH}$. One can see from this figure that the neutrino background flux from PBH evaporations is 
comparable with the theoretical curve \cite{atm} for the  
atmospheric neutrino flux at $\sim 100\text{ MeV}$ if $n\sim 1.29$ (at $T_{RH}\sim 10^{9}\text{ GeV}$).
Such large values of the spectral index are, probably, excluded by the recent large-scale experiments.

Next figure shows the constraints on the spectral index following from  data of 
neutrino experiments, as a function of the reheating temperature. As in our previous work,
we used for a description of these constraints the data of the Kamiokande atmospheric neutrino experiment \cite{25}
and the experiment on a search of an antineutrino flux from the Sun \cite{26}
(see \cite{11} for details). 

For completeness, we show on the same figure the constraints following
from extragalactic diffuse gamma ray data; part of them (NJ case) is in qualitative
agreement with the results of ref.\cite{27}.  It is seen that the constraints
are much more weak in NJ case (at $T_{RH}\agt 10^{10}\text{ GeV}$). It is again
the result of an absence of the tail of large masses in NJ spectrum, leading
to a shrinkage of the $z$-distributions and to a steepening of the background neutrino
spectra.

One  can see from Fig.\ref{fig:fig6} that, in general,
spectral index constraints following from the comparison of neutrino background predictions with 
existing data of neutrino experiments are rather weak if the purely power law of primordial density 
fluctuations is assumed for all scales (and, correspondingly, the normalization on COBE data is used).
However, it does not mean that neutrino background intensities from PBH's evaporations cannot be
noticeable. If, for example, the spectrum of primordial density fluctuations is a combination of two simple power
law spectra \cite{20}, i.e.,
\begin{eqnarray}
\label{13}
\sigma_H(M_h)\sim M_h^{\frac{1-n_s}{4}}\;,\;\;\;\;\;M_h<M_{h_c}<M_{eq}\;;\nonumber\\
\\
\sigma_H(M_h)\sim M_h^{\frac{1-n_l}{4}}\;,\;\;\;\;\;M_{eq}>M_h>M_{h_c},\nonumber
\end{eqnarray} 
then the small value of $n_l\sim 1$ ($n_l-1\approx 0$) which follows, in particular,
from COBE data, does not contradict with the possibility that $n_s-1$ is large and,
correspondingly, $\sigma_H$ at small scales is also large. If $M_{h_c}$ and $n_l$
are known from some model, one can easily obtain from the neutrino data the constraints on the 
small scale spectral index $n_s$, in the same manner as it is done above for the all 
scale spectral index $n$. Several cosmological models of such kind,  which predict 
a large density fluctuation amplitude just on small scales (while
the amplitude on large scales is constrained by the small CMBR anisotropy) 
appeared recently. The general problem of all these models is that the large
density perturbations on small  scales can lead to an overproduction of the
primordial black holes. An origin of the large density perturbations
and the scale where they are close to maximum depend on the model.
For  example, hybrid inflation models  with two stages of inflation \cite{28}
predict a characteristic spike  in the perturbation spectrum, and the model parameters (those determining
a duration  of the second stage of inflation) 
can be chosen so that PBHs evaporations give large contributions to
extragalactic backgrounds of neutrinos and $\gamma$-quanta. Inflation
models with a subsequent preheating phase also predict perturbation
spectra growing strongly toward small scales   \cite{29} (as a result
of the resonant enhancement of field fluctuations  during preheating).

At the end of this section we consider, with more details, the case
of running mass inflation models,
which also give a spectral index with rather strong scale dependence. 
The inflationary potential in these models is of the form \cite{8}:
\begin{equation}
\label{4_1}
V=V_0 \left[ 1-\frac{1}{2}c\frac{\phi^2}{Q} \right],
\end{equation}
where $c$ and $Q$ are positive constants, $\phi$ is inflation field. This form corresponds to a 
loop correction coming from softly broken global supersymmetry. The constant
$c$ is, essentially, the coupling strength of the field in the loop which is
expected to be of the order of $0.01 - 0.1$ \cite{30}. The spectral index is given by
the simple expression:
\begin{equation}
\label{4_2}
\frac{1}{2}\left[ n(k) - 1\right] = \sigma e^{-c N (k) } - c\;,
\end{equation}
where $N(k)$ is a number of $e$ - folds  of inflation that occur after
the epoch when a scale $k$ leaves horizon outside. The new parameter $\sigma$
in Eq.(\ref{4_2}) is expressed through $Q$, $c$ and $N(k)$. Using the connection
$d \ln k = -d N$ one obtains
\begin{equation}
\label{4_3}
\frac{1}{2} n'=\frac{c\sigma}{k}e^{-cN(k)}.
\end{equation}
 
For normalization of the perturbation amplitude to COBE data we use \cite{31}
the parameter $N_{COBE}=\ln (k_{end}/k_{COBE})$ connected with $N(k)$
by the expression

\begin{equation}
\label{4_4}
N(k)=N_{COBE} - \ln\frac{k}{k_{COBE}},
\end{equation}
where the COBE scale (the center of the range explored by COBE) is
\begin{equation}
\label{4_5}
k_{COBE}=\frac{7}{3000} h Mpc^{-1}.
\end{equation}

We can fix the parameter $N_{COBE}$ assuming the  instant
reheating after the end of slow-roll inflation, and expressing $N_{COBE}$ through  $T_{RH}$ using
Eq.(\ref{1_27}),
\begin{equation}
\label{4_6}
k_{fl}^{max}=(a_{eq}H_{eq})\sqrt{\frac{M_{eq}}{M_i}},
\end{equation}
the connection between $M_i$ and $T_{RH}$ [Eqs.(\ref{7}),(\ref{7a})], and the relation
\begin{equation}
\label{4_7}
k_{fl}^{max}=k_{end}=e^{N_{COBE}}a_0 H_0.
\end{equation} 
Instead of this, we put, for simplicity, $N_{COBE}=45$ (for all  possible values of $T_{RH}$), i.e., we assume that,
in a general case, $k_{fl}^{max}<k_{end}$ (radiation era begins not  immediately after the end of slow-roll inflation).
The slow-roll condition, 
\begin{equation}
\eta=\frac{1}{8\pi} M_{pl}^2 \frac{V''}{V_0}<1,
\end{equation} 
holds if  values  of the parameter $c$ are not too large.  If, e.g., $n_{COBE}=1$, this condition can
be  rewritten as
\begin{equation}
\eta\approx c e^{c  N_{COBE}} <1.
\end{equation}

The horizon  crossing amplitude $\delta_H (k)$ is calculated using Eq.(\ref{1_15}), where we put $k_0=k_{COBE}$.
The expression (\ref{4_2}) for $n(k)$ has two parameters: $\sigma$ and $c$. If we suppose
that the spectral index at $k=k_{COBE}$ is known from experiment (e.g., according to ref.\cite{32}, it follows  from
COBE  data that $n_{COBE}=1.02\pm 0.24$),   we exclude one parameter (e.g., $\sigma$).
Now we can use  the same steps as above to obtain the constraints on the  parameter $c$.
The only difference   is that now one must use the general expressions for the PBH mass spectrum, Eq.(\ref{1_20}),
and for $\sigma_H^2(M_h)$, Eq.(\ref{1_7a}).
These constraints are shown on Fig.\ref{fig:fig7} for three possible values of $n_{COBE}$. Solid lines on the figure
correspond to the critical collapse case, dotted line (calculated for the value $n_{COBE}=1.0$ only) correspond
to the standard picture of the collapse.
The lines show the upper limits  of the parameter $c$.
One can clearly see the difference between two cases: the sharp rise
of the dotted curve at $T_{RH}\sim 10^8 \text{GeV}$ is directly
connected with the sharp cut-off of the initial 
mass spectrum (for the standard collapse  case) at $M_{BH}=M_i$.
Indeed, due to the proportionality 
\begin{equation}
M_i\sim  t_i\sim T_{RH}^{-2},
\end{equation}
at small enough reheating temperatures the minimum values of the PBH mass in the spectrum  are too large
and, therefore, the large part of PBHs cannot evaporate until the present time.

Finally, on Fig.\ref{fig:fig8} we show the resulting constraint curves for the horizon crossing amplitude $\delta_H (k)$,
for fixed values of $n_{COBE}$ and $T_{RH}$ (for the critical collapse case). The curves correspond to the 
upper limits of the parameter $c$ shown on the previous figure. The end points of the power spectrum
on the figure is determined from Eq.(\ref{4_6}). It is seen from Fig.\ref{fig:fig8} that allowed values of $\delta_H^2$
at the shortest scale (for a given $T_{RH}$) are rather large, 
\begin{equation}
\delta_H^2 (k_{fl}^{max})\sim 3\cdot 10^{-4}.
\end{equation}
The corresponding $\sigma_H$ value is  $\sim 0.08$. At $T_{RH}=10^{10}\text{ GeV}$ the shortest 
scale, $k_{fl}^{max}\sim 10^{18} (h\text{ Mpc}^{-1})\sim k_{end}$,
corresponds to a rather small value of the horizon mass, $M_{h}^{min}=M_{i}\sim 10^{11}\text{g}$.
PBHs of such a small mass evaporated at $t\sim 10^6 \text{s}$.

\section{Conclusions}
\label{sec:Dis_and_Con}

\begin{figure}[!t]
\epsfig{file=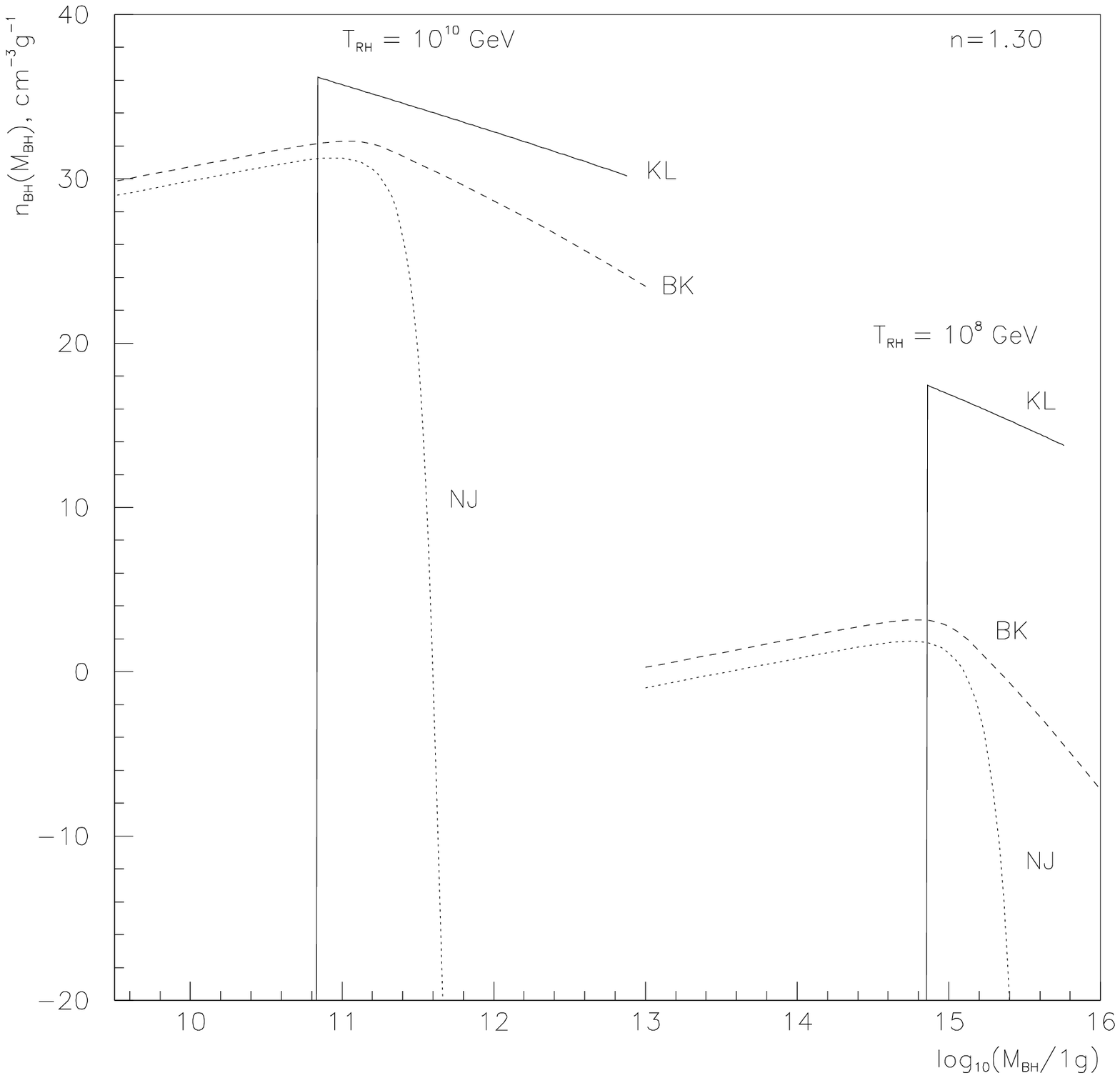,width=\columnwidth}
\caption{Examples of PBH mass spectra for two values of $T_{RH}$. Solid lines: KL spectra [18], dotted lines:
NJ spectra [16], dashed lines: BK spectra [11]; $n=1.30$ in all cases.}
\label{fig:fig1}
\end{figure}

\begin{figure}[!t]
\epsfig{file=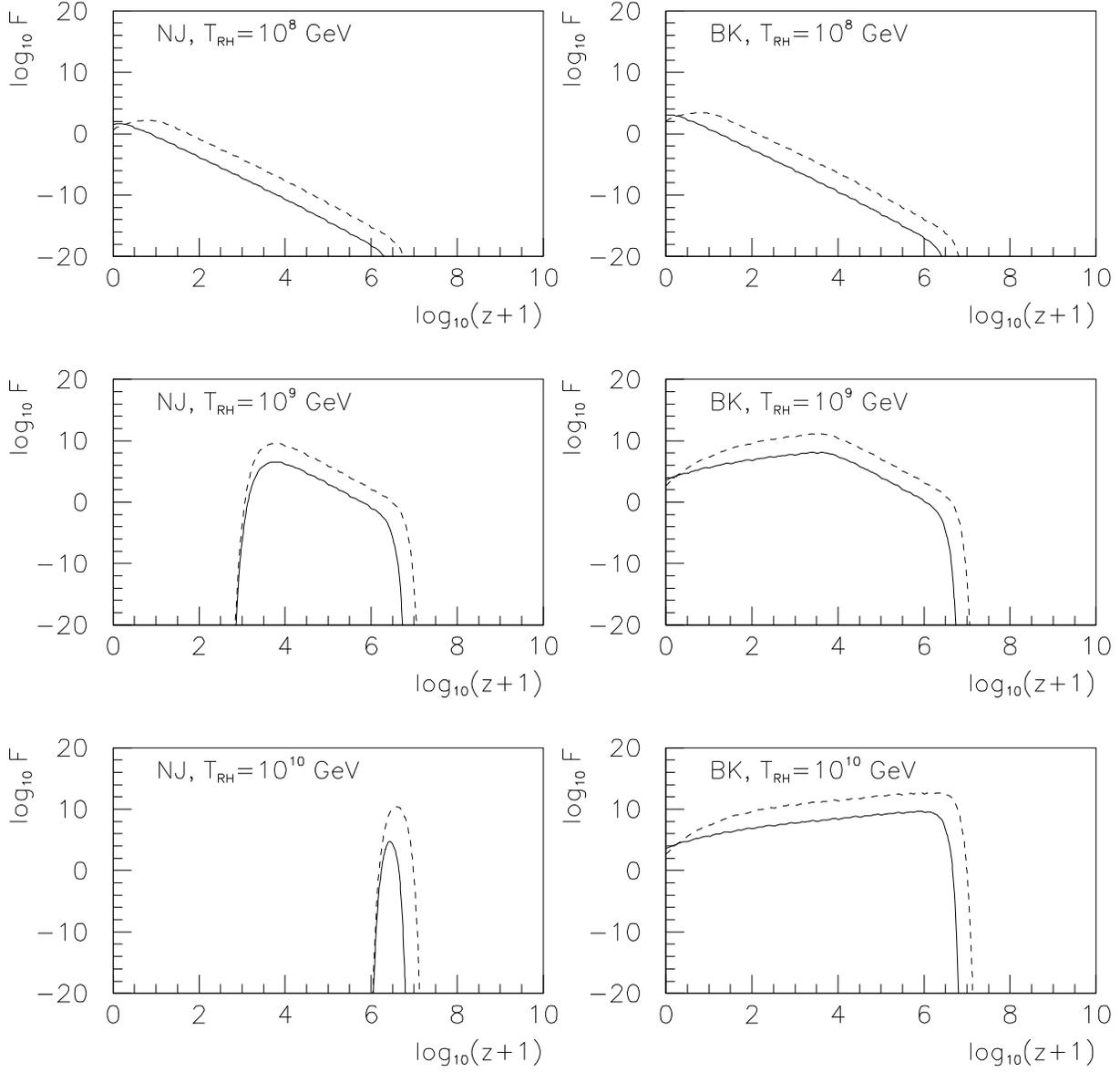,width=\columnwidth}
\caption{Red-shift distributions (integrands of the expression (\ref{10}) for the background spectrum) for neutrino energy
$E=10^{-1}\text{ GeV}$ (solid lines) and $E=10^{-2}\text{ GeV}$ (dashed lines), for PBH mass spectrum (left column) and
BK mass spectrum (right column), for three values of $T_{RH}$; $n=1.30$ in all cases. The dimension of $F(z)$ is ($s^{-1} cm^{-2} sr^{-1} GeV^{-1}$).}
\label{fig:fig2}
\end{figure}

\begin{figure}[!t]
\epsfig{file=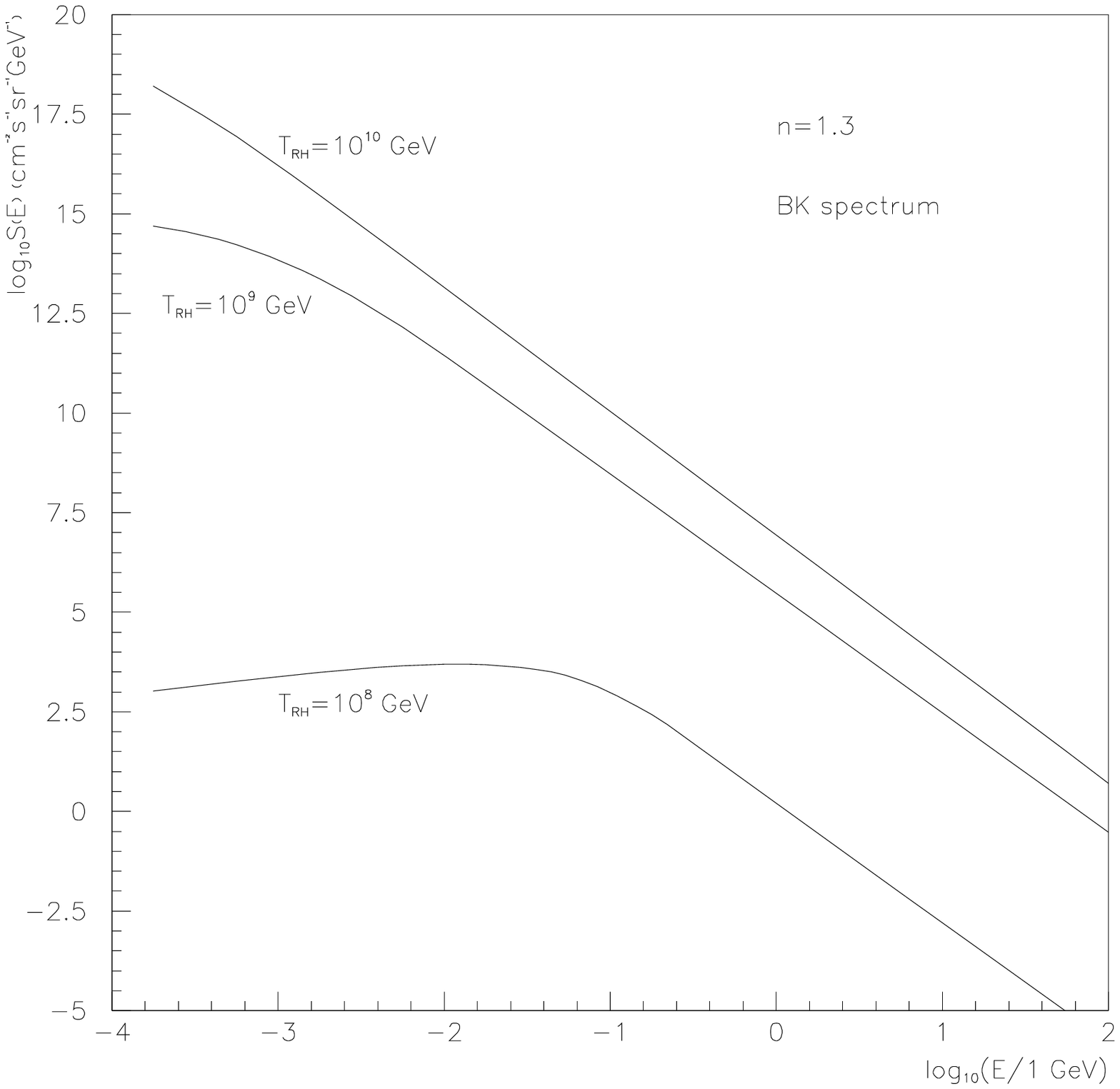,width=\columnwidth}
\caption{Neutrino background spectra for three values of $T_{RH}$, for BK mass spectrum; $n=1.30$.}
\label{fig:fig3}
\end{figure}

\begin{figure}[!t]
\epsfig{file=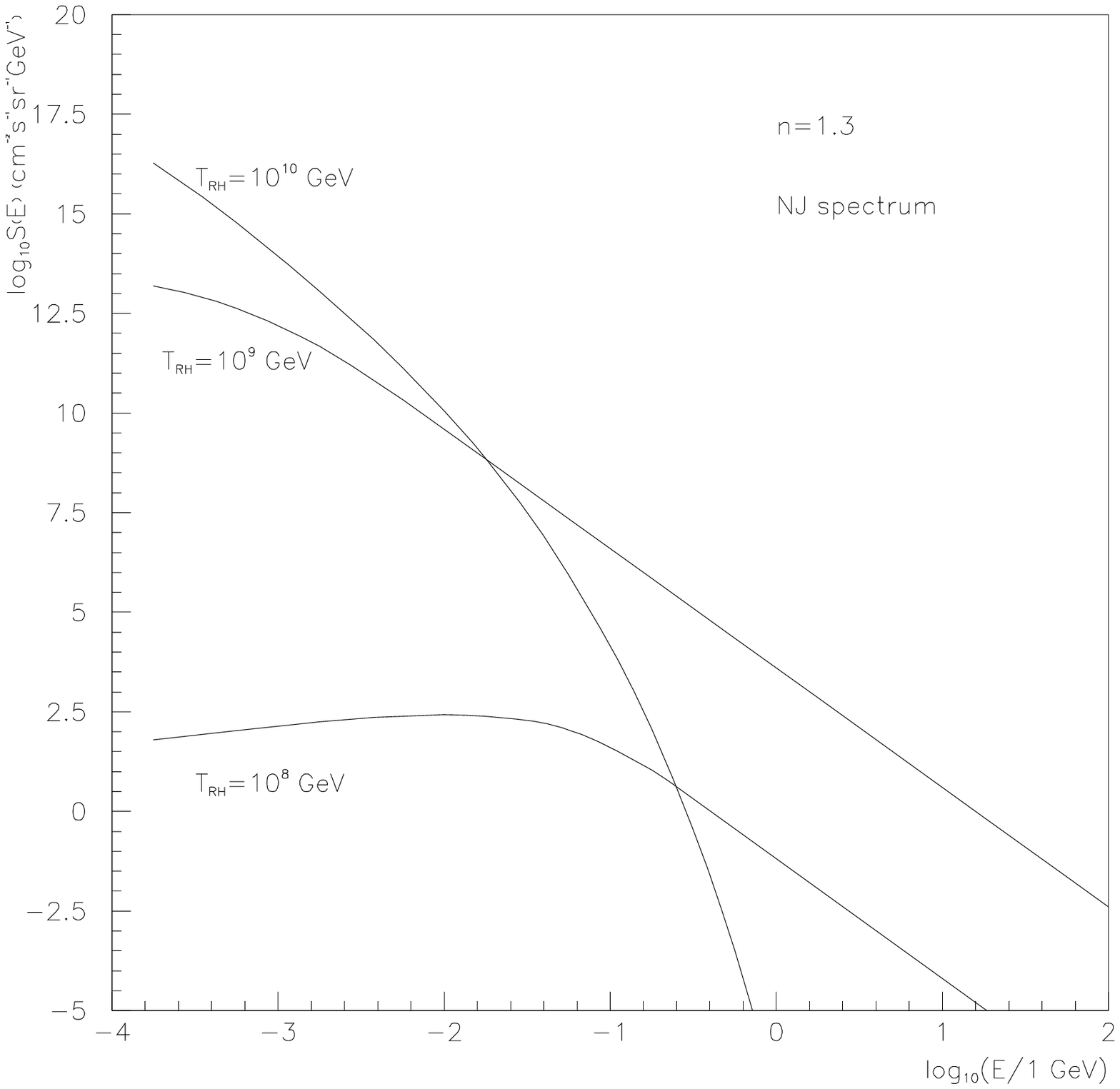,width=\columnwidth}
\caption{Neutrino background spectra for three values of $T_{RH}$, for NJ mass spectrum; $n=1.30$.}
\label{fig:fig4}
\end{figure}

\begin{figure}[!t]
\epsfig{file=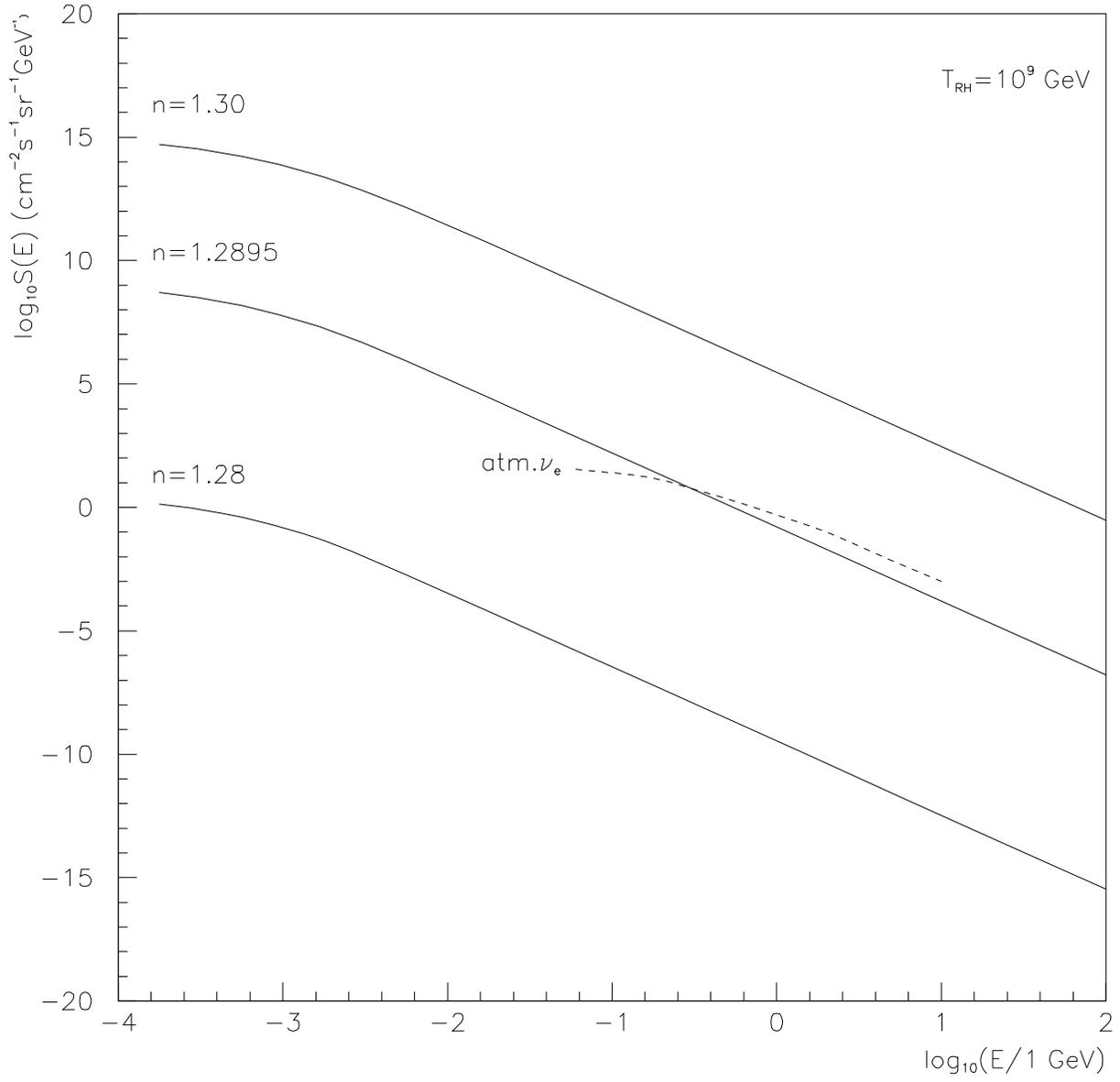,width=\columnwidth}
\caption{Neutrino background spectra for three values of the parameter $n$, for BK mass spectrum. Dashed line
represents the theoretical atmospheric neutrino spectrum for Kamiokande site (averaged over all directions) [25].}
\label{fig:fig5}
\end{figure}

\begin{figure}[!t]
\epsfig{file=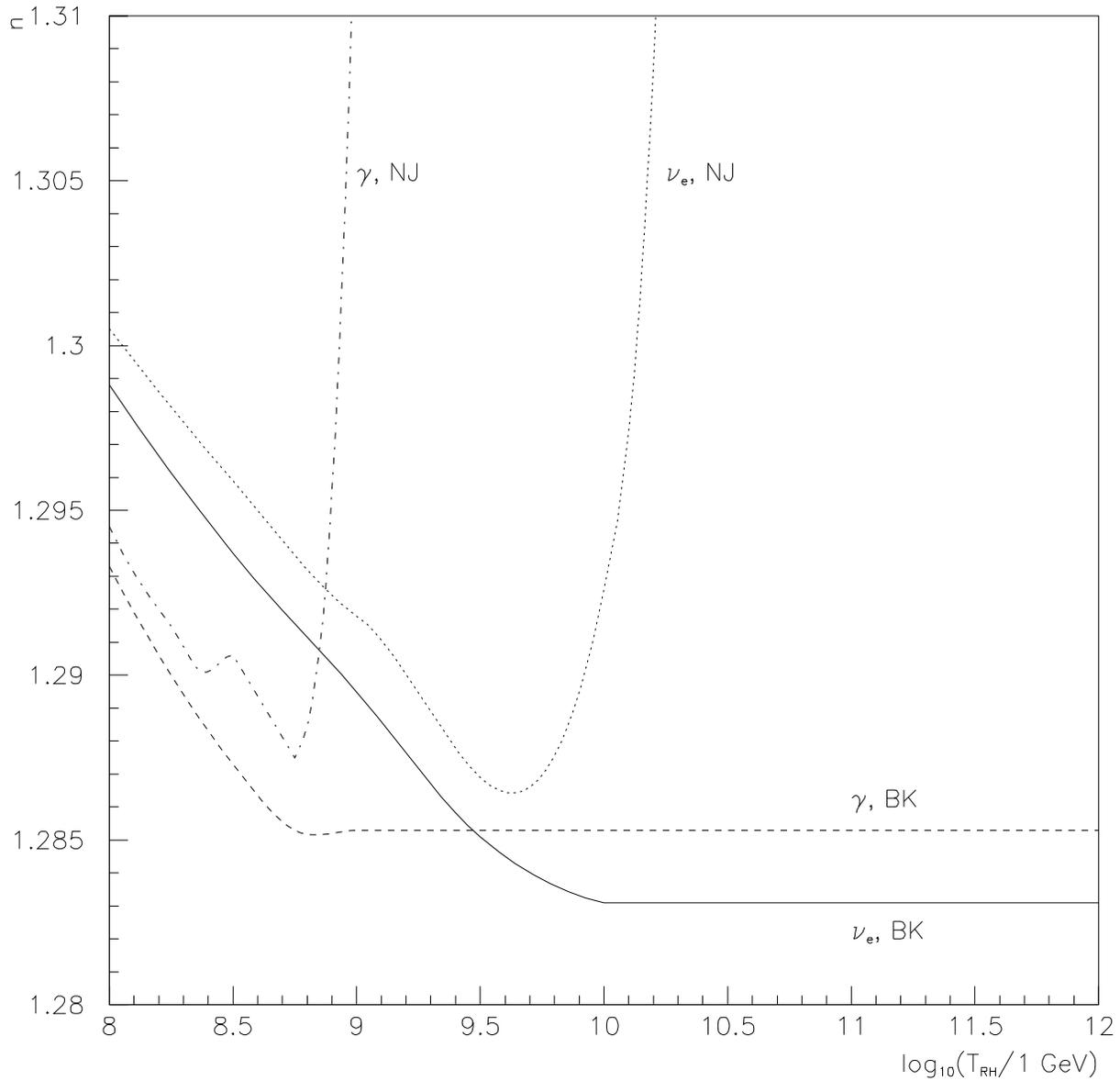,width=\columnwidth}
\caption{Constraints on the spectral index $n$ as a function of the reheating temperature $T_{RH}$. Solid and dotted lines:
BK and NJ mass spectra, atmospheric $\nu_e$ and solar $\tilde \nu_e$ experiments. Dashed and dot-dashed lines: BK and NJ mass
spectra, extragalactic diffuse gamma-ray background data.}
\label{fig:fig6}
\end{figure}

\begin{figure}[!t]
\epsfig{file=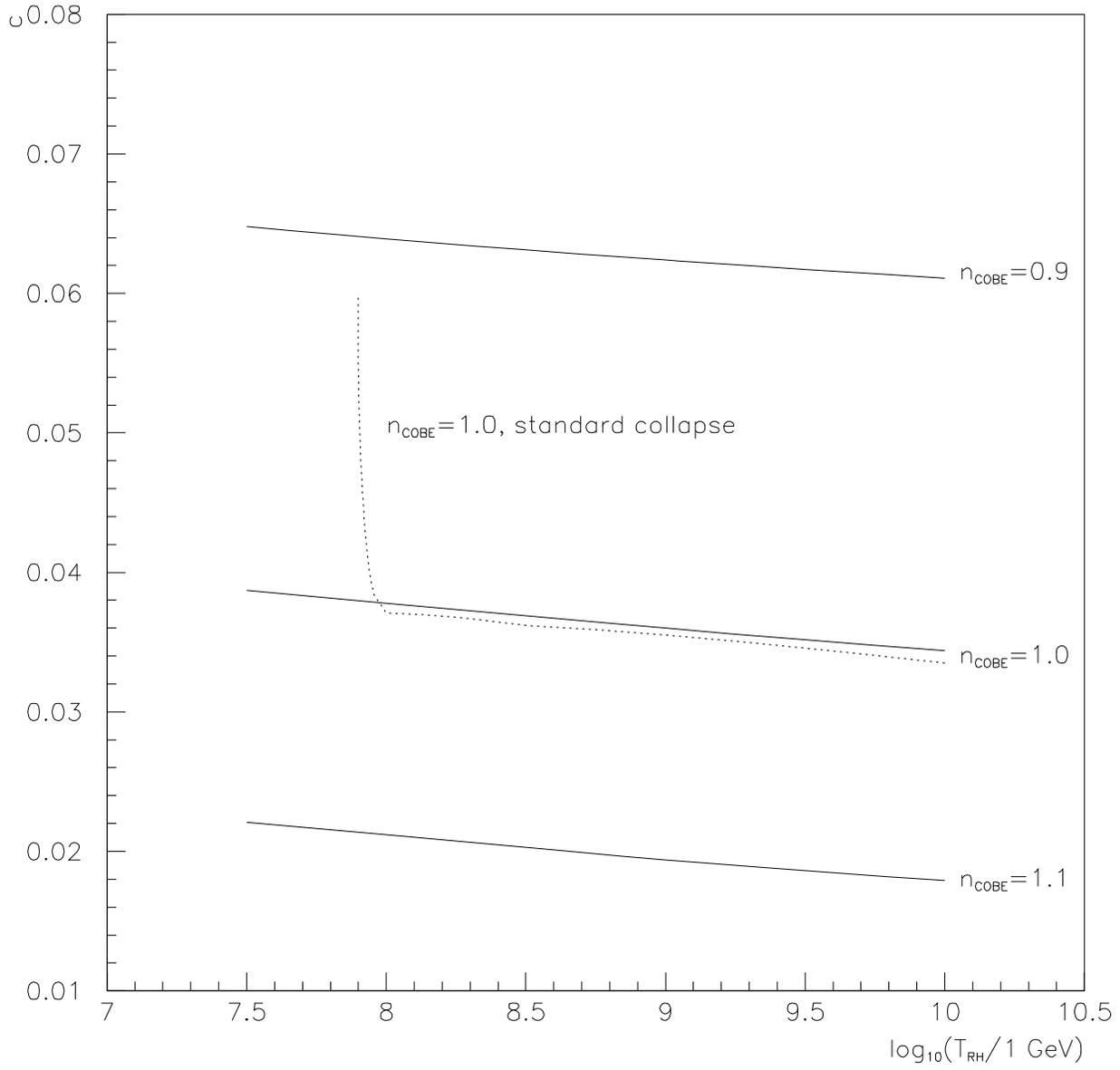,width=\columnwidth}
\caption{Upper limits on values of the parameter  $c$ of  the running mass inflation model, for different reheating temperatures and for
several possible values  of $n_{COBE}$. Solid lines: critical collapse case, dotted line: standard collapse case.}
\label{fig:fig7}
\end{figure}

The main conclusion of  the work is the following:
constraints on parameters of cosmological models from evaporations of primordial black holes
depend rather  strongly on a form of their initial mass spectrum. Number densities of PBHs predicted by the models  are extremely
sensitive to values of primordial density perturbation amplitudes and, therefore, even very steeply falling
mass spectra of PBHs can give useful constraints.
As it follows from Fig.\ref{fig:fig6}, constraints on the scale independent spectral index $n$ at high
values of a reheating temperature are drastically different  in NJ
and BK cases. We thoroughly traced the origin of this result: the absence of a  long {\it high  mass tail} in the
initial NJ spectrum (Fig.\ref{fig:fig1}) leads to a relative shrinkage of the redshift distributions  of  
evaporated neutrinos (Fig.\ref{fig:fig2}) and,  as a consequence, to a steepening of the neutrino background spectra and 
to a decrease of the background intensities (Fig.\ref{fig:fig4}),
resulting, eventually, in a weakening of the spectral index constraint.
So, the constraint values at $T_{RH}> 10^{10}\text{GeV}$ on Fig.\ref{fig:fig6} are entirely 
due to  an effect of the summation over all epochs  of the
PBH  formation in the approach  of ref.\cite{20,21}.
Analogously,  we see,  on an example of the parameter $c$ of running mass inflation models, that the constraints  at low
values of $T_{RH}$ also depend  on a type of the gravitational collapse: in the standard collapse case
the constraint at $T_{RH}< 10^{8}$ is practically absent (dotted  curve on Fig.\ref{fig:fig7})
due  to a corresponding  absence of a {\it low mass tail} in the initial
PBH mass spectrum of this case (compare KL and BK spectra on Fig.\ref{fig:fig1}; the behavior of the corresponding PBH mass spectra in a case of 
a scale dependent $n$ is qualitatively similar).

The initial number density of PBHs is larger in a case of the standard picture of
the gravitational collapse (as compared with the critical collapse case). It is due to
relatively small value of the critical overdensity ($\delta_c^{st}=\gamma=1/3$). In the 
region near the maximum of PBH mass spectra the ratio of intensities is given by the 
approximate relation
\begin{equation}
\label{12}
\frac{n_{BH}^{(KL)}}{n_{BH}^{(BK)}}\sim e^{\frac{\delta_c^2-\gamma^2}{2\sigma_H^2}}\;.
\end{equation}
Correspondingly, intensities of the neutrino and gamma backgrounds produced by PBH's evaporations
are smaller and spectral index constraints are systematically weaker in the critical collapse case
(see, for  the comparison with  the standard collapse case, Fig.10 of ref.\cite{11}).

It is worth noting that spectral index constraints followed from neutrino evaporations are stronger at high $T_{RH}$
values (as compared with the constraints from $\gamma$-quanta), especially
in the NJ case (Fig.\ref{fig:fig6}). The reason is simple:
at high $T_{RH}$ the relative contribution of large redshifts in $z$-distributions
is very large (Fig.\ref{fig:fig2}) and, correspondingly, the absorption of $\nu$,~$\gamma$ during
propagation in space is important. Naturally, absorption effects
are more sufficient for $\gamma$-quanta than for neutrinos (for $\gamma$-quanta $z_{max}\sim 700$, while for neutrinos
$z_{max}\sim 10^7$ \cite{11}, as we can see, in particular, from curves  of Fig.\ref{fig:fig2}).

\begin{figure}[!t]
\epsfig{file=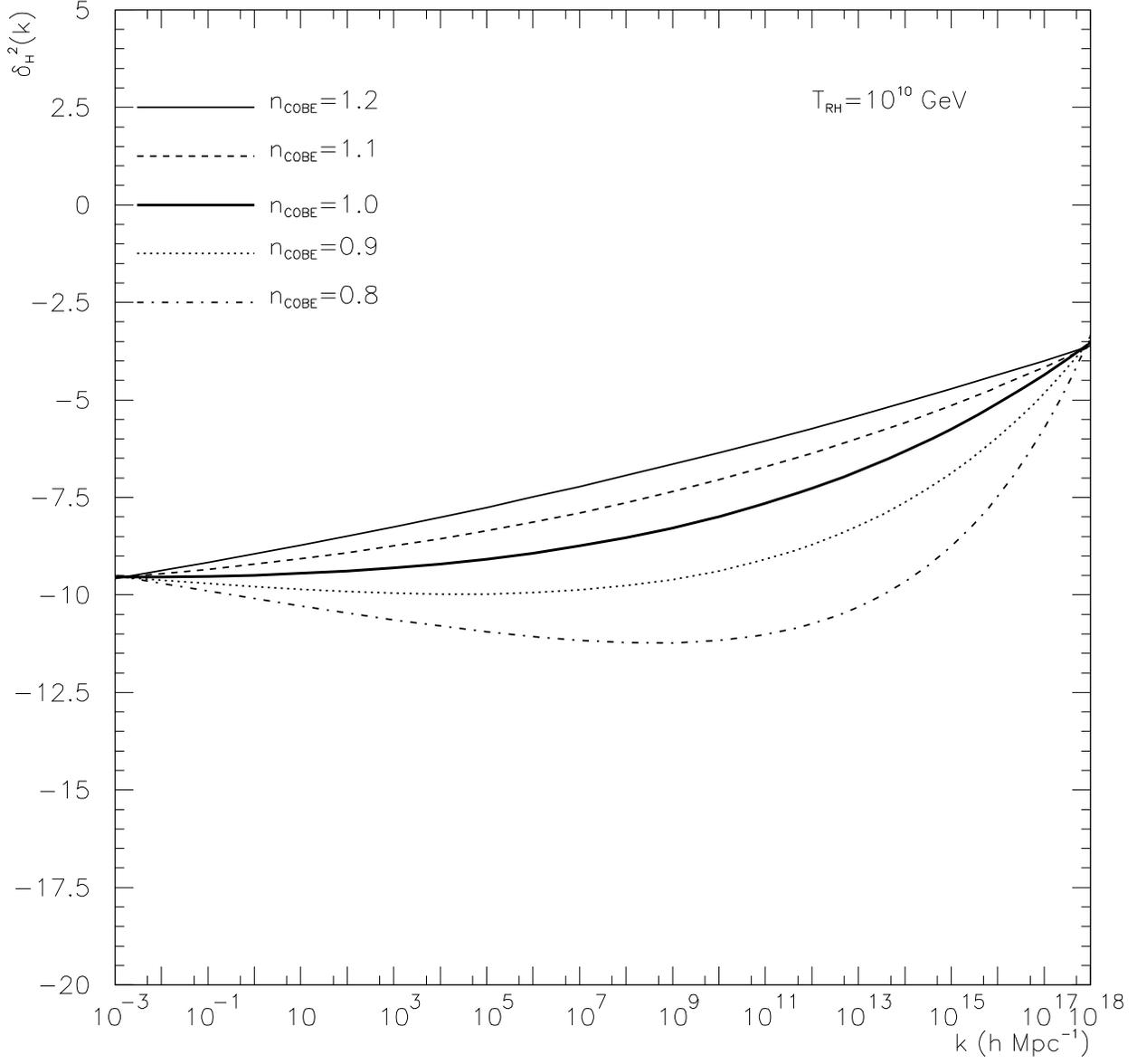,width=\columnwidth}
\caption{Constraints on the power spectrum of primordial density fluctuations in the running mass inflation model, 
for several values of $n_{COBE}$ and for $T_{RH}=10^{10}\text{GeV}$.
All curves are calculated for the critical collapse case.}
\label{fig:fig8}
\end{figure}

One should  note that, in general, the constraints on a scale independent spectral  index  
followed from  PBH evaporations (Fig.\ref{fig:fig6})  are  rather  weak. Probably,
so large deviations of the spectral index from 1 ($n\sim 1.28$) are excluded by the latest data (see,  e.g., \cite{30}). 
Therefore, constraints from PBH evaporations  are  more useful for those  cosmological models
in which the spectral  index is  scale dependent. In such models large density perturbation amplitude at small scales  
(and, correspondingly, large effects from evaporations) can  coexist with the small value of    $\delta_H (k)$ at COBE scale   
(Fig.\ref{fig:fig8}). We  showed in this paper,
using, as an example, the running mass inflation  model, that in this case  the PBH evaporation  process can  be
used for  a  constraining of model parameters [in particular, parameters of an inflationary potential, (Fig\ref{fig:fig7})].
We stress, once more, that all constraints of such type depend on assumptions used in a derivation
of the initial PBH mass spectrum.

\end{document}